\documentclass{elsart}

\usepackage{natbib}

 \usepackage{graphics}

\begin{document}

\begin{frontmatter}



\title{Properties of Dark Matter Haloes}


\author{F. Combes}

\address{Observatoire de Paris, LERMA, 61 Av. de l'Observatoire,
F-75014, Paris, France}

\begin{abstract}
An overview is presented of the main properties of
dark matter haloes, as we know them from observations,
essentially from rotation curves around spiral and
dwarf galaxies. Detailed rotation curves are now known for
more than a thousand galaxies, revealing that they are
not so flat in the outer parts, but rising for late-types,
and falling for early-types.  A well established result now
is that most bright galaxies are not dominated by dark
matter inside their optical disks. Only for dwarfs and LSB 
(Low Surface Brightness galaxies) dark matter plays
a dominant role in the visible regions. The 3D-shape of haloes are
investigated through several methods, that will be discussed:
polar rings, flaring of HI planes, X-ray isophotes.
 It is not yet possible with rotation curves to know
how far haloes extend, but tentatives have been made.
It will be shown that the dark matter appears to be coupled
to the gas in spirals and dwarfs, suggesting that dark
baryons could play the major role in rotation curves.
Theories proposing to replace the non-baryonic dark matter by a different
dynamical or gravity law, such as MOND, have to take into account
the dark baryons, especially since their spatial distribution
is likely to be quite different from the visible matter.
\end{abstract}


\end{frontmatter}

\section{Rotation Curves}
\label{rotation}

At galactic scales, the best tools to probe the dark matter content
of the universe are combined HI and H$\alpha$ or CO 
rotation curves of spiral galaxies
(see the review by Sofue \& Rubin 2001). The optical rotation curves
provide high spatial resolution in the visible disk, and in particular
in the center, to trace central mass concentrations, while only the HI
gas extend far enough in radius to trace the outer parts, where dark
matter is dominating. A lot of progress has been made recently in our knowledge
of dark matter content of galaxies, because of large samples observed
in 2D Fabry-Perot H$\alpha$ spectroscopy, and also I-band or near-infrared
photometry (Mathewson et al 1992, Schommer et al 1993, Eskridge et al 2000). 
B-band images of
galaxies suffer from extinction, in particular in the center of galaxies,
leading to underestimating the stellar disk contribution to the mass,
and magnifying the contribution of an hypothetic dark component. Also 
the mass-to-light ratios are varying more strongly with stellar
populations in the blue. This is illustrated by the larger scatter of
the Tully-Fischer relation in the blue (e.g. Verheijen 2001).

\begin{figure}
\resizebox{12cm}{!}{
{\includegraphics{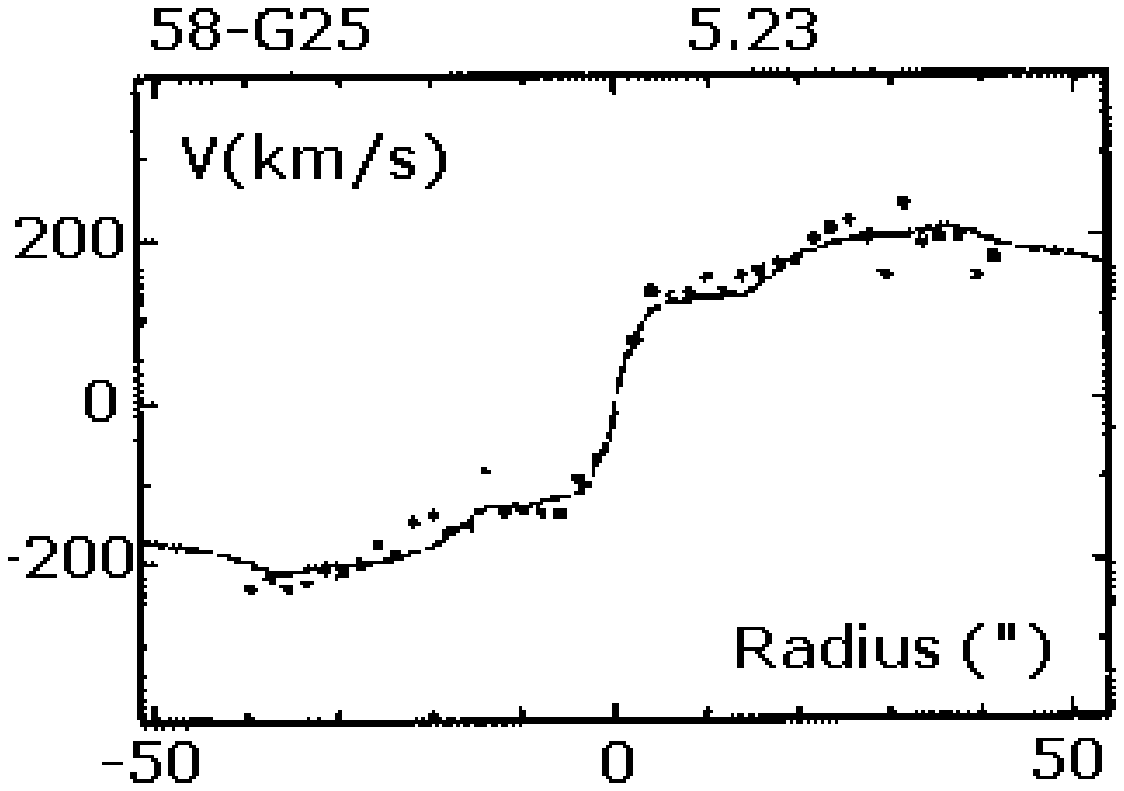}} 
{\includegraphics{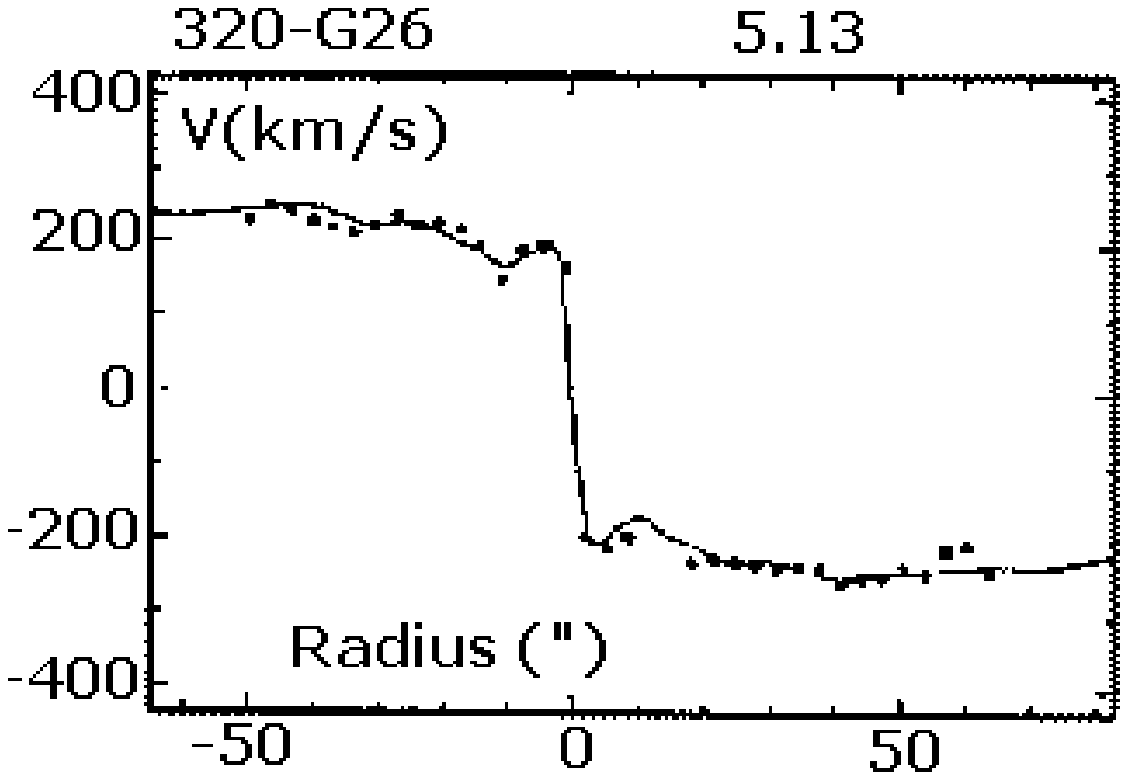}}  }
\caption{ Examples of H$\alpha$ rotation curves (dots) and their
fits with I-band images (full lines); the 
corresponding M/L ratios is indicated above each panel,
from Buchhorn (1993).}
\label{buchhorn}
\end{figure}

\subsection{Rotation curves of normal spirals}
 
The new feature resulting from these recent surveys is that for most
spiral galaxies, the dark matter is not dominant within the optical disk.
Indeed, the 500 rotation curves observed by Mathewson et al. (1992)
have been reproduced remarquably well by Buchhorn with mass-to-light 
ratios constant with radius (e.g. Freeman 1993, and fig \ref{buchhorn})
and with values compatible with what is known from stellar populations.
 The fact that the baryonic matter is actually dominant is reflected
by the very good fit of all oscillations or "wiggles" in the observed rotation curves,
corresponding to spiral arms in the disk. A non-baryonic component would
not follow the spiral instabilities in the disk, and would have diluted
these oscillations in the rotation curves.

This point is related to the maximum disk hypothesis: the latter tries to 
fit rotation curves in attributing the maximum mass to the disk, compatible
to the central part of the curve. Then, keeping the M/L ratio constant with radius,
the rotation curve happens to be reproduced quite well over the optical disk,
without dark matter. Of course, it is still possible to reduce M/L of the 
stellar component, and also fit the rotation curve with the addition of
a dark matter component. But the peculiar streaming motions features are
then less well reproduced (Sackett 1997, Palunas \& Williams 2000).
 
Also, the fact that most galaxy disks possess bars, and these bars
are rotating rapidly (their cororation is located through resonances
in the middle of the disk), pleades in favor of a disk dominated by
the visible matter, with a negligible contribution of spherical
dark matter;  dynamical friction against the dark matter component
would slow down the bars in a few dynamical times (Debattista \& Sellwood
1998). 

\begin{figure}
\resizebox{6.8cm}{!}{\includegraphics{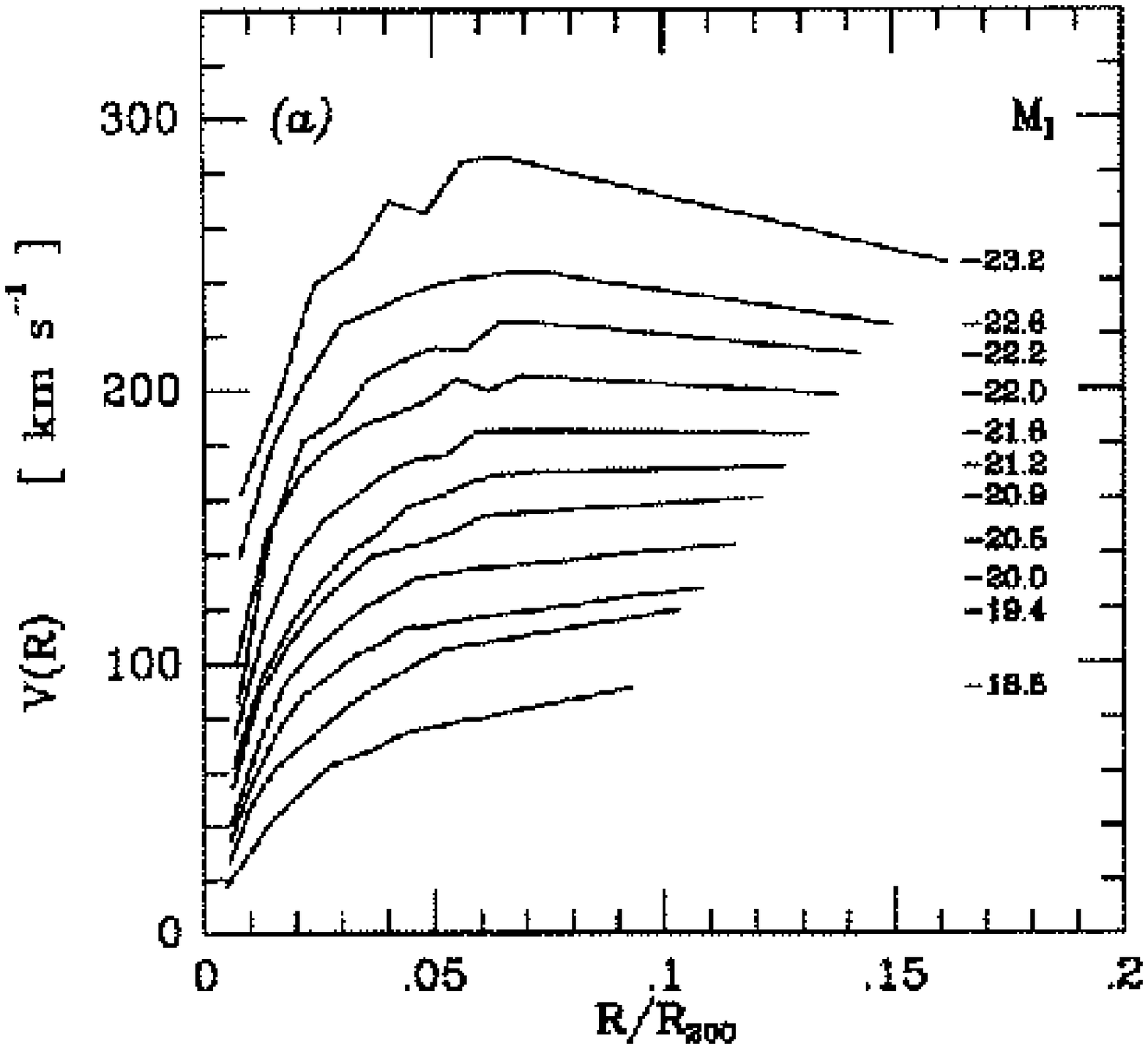}}
\resizebox{5.2cm}{!}{\includegraphics{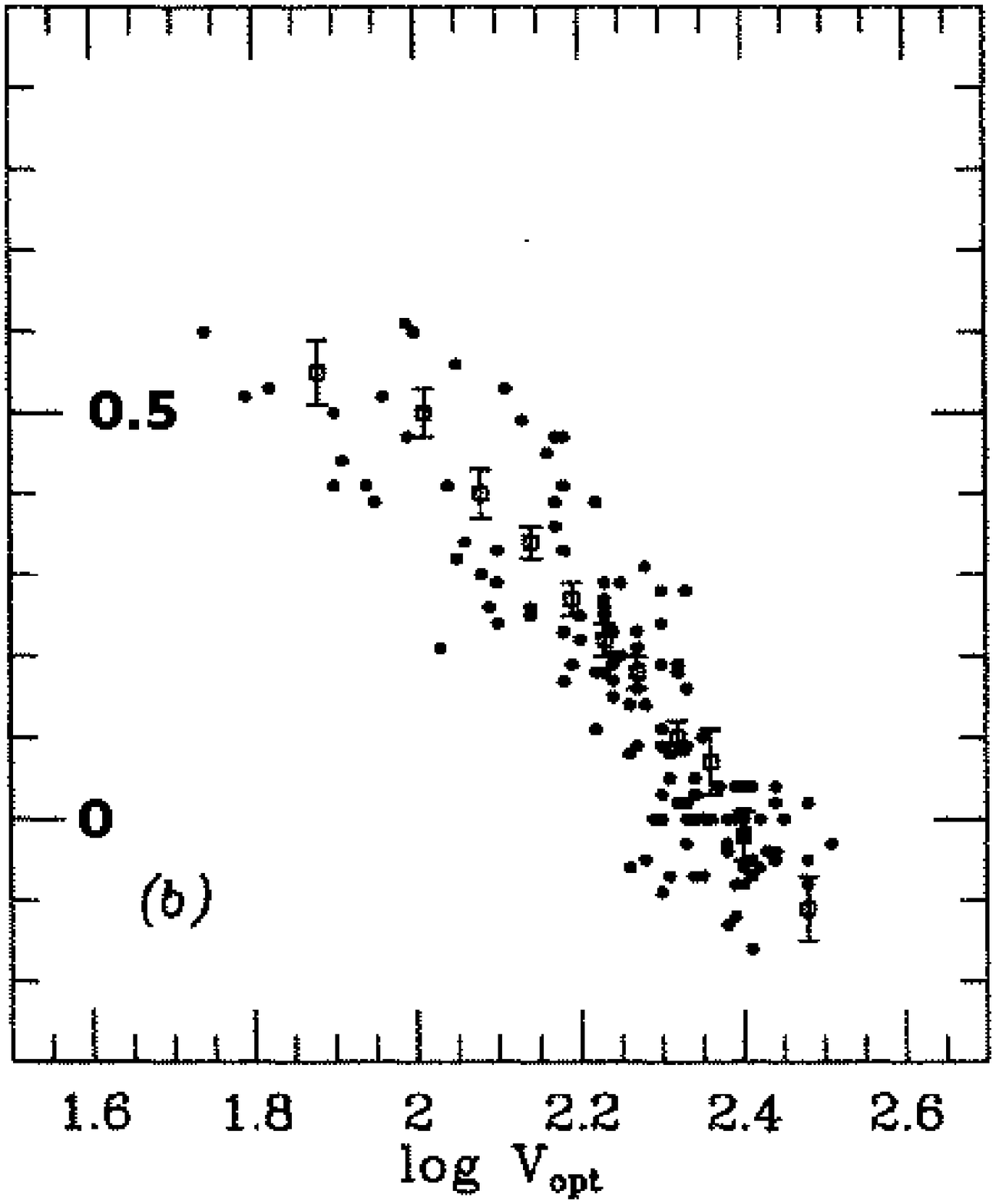}}
\caption{ {\bf (a)} The universal rotation curve of spiral disks at different
luminosities (M$_I$); radii are normalised to R$_{200}$, the mean radius 
containing a mean halo overdensity of 200.  {\bf (b)} The slope of the rotation
curve in the region (0.6-1) R$_{opt}$ versus the rotation velocity V$_{opt}$ at  
R$_{opt}$. From Persic et al. 1996). }
\label{persic96}
\end{figure}

The bottom line is therefore that dark matter is only needed 
at large radii, in the HI-21cm extensions. The fact that the 
rotation curve is flat in the outer parts, while it is no longer 
attributable to the stellar disk or bulge, has been called the
conspiracy. Why does the velocity due to the spherical 
non-baryonic dark matter coincide exactly to that of the 
stellar component?

In fact, rotation curves are not all flat, depending on their 
morphological types (Casertano \& van Gorkom 1991): early-type galaxies
have rotation curves that begin to fall down, while late-type and dwarfs
have not yet reached their maximum velocity at the last observed radius.

When compiling 1100 rotation curves, and normalising them to
their exponential radial scales, Persic et al. (1996) found that spiral 
discs (once the contribution of the bulge is removed in galaxies), may have
an universal rotation curve, only determined from their total luminosity
(cf fig \ref{persic96}).
At high luminosities, there is no or only a slight
discrepancy between the observed rotation curve and that contributed
by the luminous matter, while a larger dark matter component is required
at low luminosities: the dark-to-luminous mass ratio scales inversely with luminosity
(fig \ref{persic96}).
The halo core radius is comparable to the optical radius, but shrinks for low luminosities.

However, to draw these conclusions, Persic et al. (1996)  assumed
that there is a constant ratio
between the end-radius of the visible disk (R$_{23.5}$), and the exponential 
characterictic radius, or in other words, that all disks have the same shape.
This is not quite true, as emphasized by Palunas \& Williams (2000). the latter authors
have carried out a detailed study of 74 galaxies, where 2D Fabry-Perot
H$\alpha$ spectroscopy exist (Schommer et al 1993) and I-band photometry. Very good
fits of the rotation curves are obtained without dark matter, out to R$_{23.5}$, with 
a constant M/L. They conclude that mass traces light, in particular since the surface
brightness profiles of the various galaxies present pronounced differences. The small number
of galaxies with a poor fit have strong non-axisymmetric structures (bars and strong
spiral arms). The resulting I-band M/L = 2.4 $\pm$ 0.9 h$_{75}$, is compatible with normal
stellar populations. This indicates that the dark matter is not dominant within optical
disks, or is perfectly coupled to the visible matter.

Already this fact is contradictory to expectations from
CDM scenarios. CDM halo profiles are centrally concentrated, and
numerical simulations predict that the dark matter dominates inside
spiral disks. For example in a galaxy of the mass of 
the Milky Way, $\Lambda$CDM simulations
predict three times more dark matter than is observed
(Steinmetz \& Navarro 2000).
On the contrary, this fact is in agreement with MOND hypothesis.

\subsection{Rotation curves of dwarfs and LSB}

The relative importance of dark matter is increasing towards late types
and dwarf irregular galaxies are completely dominated by dark matter.
They are ideal tools to probe theories of dark matter, since the
uncertainties on the stellar mass-to-light ratio has negligible influence on the derived 
radial matter profile. For the prototype of these dwarfs, DD0154, the rotation curve is 
well determined until 15 optical scale lengths; the HI gas component 
is more massive than the stellar
 disk (Carignan \& Beaulieu 1989).

The derived radial profile of dark matter in dwarfs is not peaked towards the center, since 
the rotation curves are slowly rising. This is one of the main problems for the $\Lambda$CDM 
theories: the radial distribution is predicted by simulations to be highly peaked, with a cusp,
 or density following a power-law of slope -$\alpha$ = -1.5 (Navarro, Frenk \& White 1997, 
Fukushige \& Makino 97). Observed rotation curves points towards no cusp, but cores (Moore 
1994, Dalcanton \& Bernstein 2000). According to Burkert \& Silk (1997), this problem could 
ony be solved by the introduction of baryonic dark matter inside the optical disk, with a mass 
several times the visible mass, and with a similar radial distribution.

However, there are still uncertainties in the mass-to-light ratios, and the rotation curves 
are not fully sampled in all dwarf galaxies available, so that it might be still difficult to 
conclude for all of them (Swaters 1999, van den Bosch \& Swaters 2000). New models of dark 
matter have been proposed to solve precisely this problem, self-interacting dark matter with 
a non-zero cross-section (Spergel \& Steinhardt 2000), but many new problems then appear. 
Other mechanisms have been proposed, such as stellar feedback, to reduce central densities of
 CDM (Navarro et al 1996, Binney et al 2001); but this mechanism has very low efficiency, as
 soon as the galaxy is more massive than 10$^7$ M$_\odot$ (Mc Low \& Ferrara 1999).

Low Surface Brightness galaxies (LSB) are also dominated by dark matter; they can be dwarfs, 
but also massive galaxies, with a large amount of HI gas. Their rotation curves are also good
 constraints for dark matter models. Again, they are incompatible with the cuspy profiles 
predicted for $\Lambda$CDM, but can be fitted with models where matter follows light, although with too
large mass-to-light ratios (de Blok et al 2001).

\section{3D-shape of Haloes}
\label{shape}

For the sake of simplicity, many models choose a spherical shape for the dark matter component, 
but this particular shape is very unlikely.  All current scenarios predict in fact flattened 
shapes, more or less flattened according to dark matter candidates.

\subsection{Axis ratio in the galactic plane}

CDM simulations end up with triaxial shapes for collapsed structures, so that the haloes are 
not axisymmetric even in the plane of the baryonic galactic disk. This can be checked through 
the orbits of the baryons, and in particular the HI gas, with low velocity dispersion. Of course,
 inclination effects have to be taken into account, as well as flaring, warps or other distortions,
 due to the spiral, bars or ring features in the galaxy disks. 

The result of these investigations is that galaxies are actually
axisymmetric in their planes, with a very low upper limit for the excentricity: below 0.1 with the
 isophote shape versus
HI velocity widths method (Merrifield 2002), or even less than 0.045, when using near-infrared data
 to avoid extinction (Rix \& Zaritsky 1995). On special cases, the limit can be better, excentricity
 of the order of 0.012 in potential in the very regular early-type galaxy IC2006 with an HI ring
 (Franx et al 1994).
 
This axisymmetric shape of galactic haloes is confirmed by the low scatter observed for the 
Tully-Fisher relation.

\subsection{Axis ratio perpendicular to the plane}

The flattening in the direction perpendicular to the galactic plane is more difficult to establish.
 Predictions are slightly different according to the nature of dark matter.
Non-baryonic pure CDM haloes are predicted already quite flattened in numerical simulations; they are
 half oblate and half prolate,
with axis ratios of the order of c/a = 0.5, b/a = 0.7. It is interesting to note that the dark haloes
 are predicted more flattend then observed elliptical galaxies;
the distribution peaks at E5 (while elliptical galaxies
peak at E2 (c/a = 0.8)!) cf Dubinski \& Carlberg (1991).

However, these predictions of pure dark matter simulations were already incompatible with the
 observed axisymmetry of galaxy disks described above. The dissipative infall of gas in
 non-baryonic dark haloes should be taken into account. This 
concentrates even more the haloes, through adiabatic contraction, and also forces them to an 
oblate shape, and the prediction now become in average: c/a = 0.5, b/a $>$ 0.7
(Katz \& Gunn 1991; Dubinski 1994). 

As for the self-interacting dark matter model (SIDM), the predicted shape is almost spherical.
Bullock (2002) reconsider $\Lambda$CDM simulations and found rounder haloes; the warm dark matter
 models predict even more spherical haloes.

If the dominant dark matter around spiral galaxies is baryonic, and in the form of cold gas, it
 will be dissipative, and is predicted more flattened (Pfenniger \& Combes 1994). Account must 
be taken however of the strong flaring of disks in the outer parts, that makes the potential 
rounder. The very frequent warps of galaxy planes in the outer parts, related to gas accretion 
and long relaxation time there, also accentuate the roundness of the potential.

\subsection{Polar rings}
Polar ring galaxies are peculiar objects composed of two systems with almost perpendicular angular
 momenta: a host galaxy, early-type in general (a lenticular more frequently), is surrounded by 
a perpendicular ring of gas and stars following nearly polar orbits.
These systems are thought to be formed during an accretion or merger event. They are quite 
precious tools to probe the 3D shape of dark matter haloes, since HI gas is orbiting perpendicular
 to the main plane of the host galaxy.

Many problems however have prevented a clear picture to emerge:
\begin{itemize} 
\item the early-type host system has a stellar component with large velocity dispersion (certainly
 heated by the accretion event), and the derivation of its rotation curve is model dependent,
  
\item gas cannot coexist at the same radius in perpendicular planes, since collisions will
dissipate its energy quickly, and it will infall to the center. Therefore, it is impossible
to compare equatorial and polar velocities at a given radius (and indirect comparisons at different
radii are model dependent,) 

\item due to an obvious selection effect, observed polar rings are massive, and
therefore the polar matter cannot be considered as test particles to probe 
the host galaxy potential, but the polar mass perturbs the potential, and the polar
ring might even be self-gravitating.
\end{itemize} 

The estimation of the potential flattening around one of the best known 
polar ring galaxy NGC4650A has given rise to very different results: either 
spherical, or flattened along the host galaxy plane, or flattened along
the polar ring itself (Sackett et al. 1994, Combes \& Arnaboldi 1996).
The latter geometry could be explained in the case of a non perturbative 
merger, where a massive galaxy settles with its flattened dark halo,
perpendicular to the host lenticular system (Bekki, 1998), 
or in case of gas accretion, if gas is representing a significant part
of the dark matter around galaxies (Bournaud \& Combes 2002). The fact that
polar ring galaxies have a wider HI velocity width with respect to corresponding
galaxies in the Tully-Fisher relation (Iodice et al 2002) supports this
hypothesis. 

\begin{figure}
\rotatebox{-90}{
\resizebox{7cm}{!}{\includegraphics{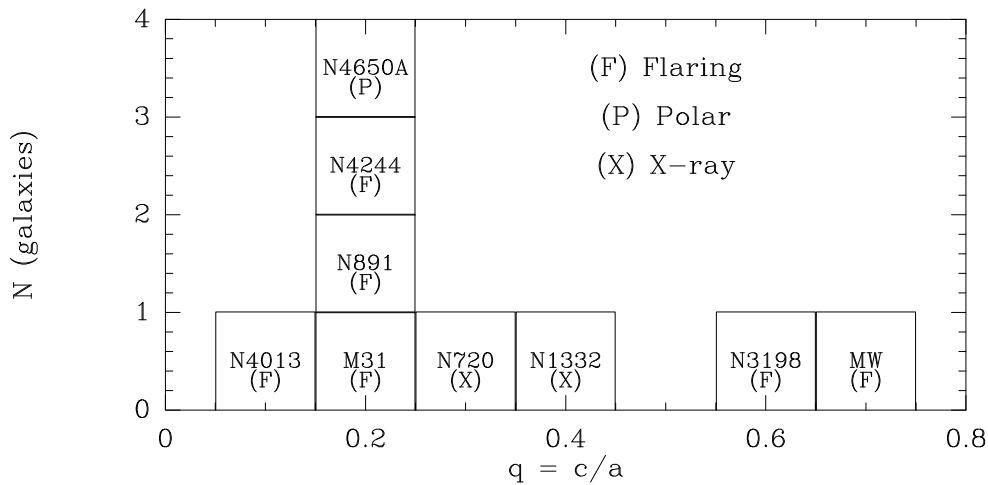}}}
\caption{Histogram of halo flattening, according to 
the three methods used in the litterature:
flaring gas-layer, polar-rings and X-ray isophotes}
\label{flat}
\end{figure}

\subsection{Flaring of the HI plane}

In the outer parts of galaxies, if the gas is in hydrostatic equilibrium,
its scale height is a function of the density of dark matter 
in the plane (which provides the restoring force towards the plane), and of the gas
velocity dispersion. The HI gas dispersion is observed to be almost constant
with radius in face-on galaxies: 
$\sigma_z$ (HI) = 10 km/s  (Dickey et al 1990, Kamphuis, 1992), and since the surface density
of dark matter is falling with radius (to provide the flat rotation curve),
the restoring force is declining, and therefore the gas is flaring with radius.
The amount of flaring is an indicator of the surface density of dark matter in
the plane, with respect to a spherical distribution which produces only radial
forces.  The method was applied to the galaxy
NGC 4244 by Olling (1995, 1996), and to NGC 891 by Becquaert \& Combes (1997),
and both haloes were found quite flattened, with an axis ratio of
q= c/a= 0.2. But here too, some free parameters have to be chosen,
that can moderate these results. First, the flaring of the HI plane is
well measured for edge-on galaxies, but then the 
velocity dispersion perpendicular to the plane is not well known,
and must be extrapolated from other galaxies, or derived from a model.
 More important, the deduced degree of flattening depends
on the assumed truncation radius of the dark matter halo. 
Indeed, contrary to a spherical distribution, the forces inside a certain
radius is strongly dependent on the mass outside, if the halo is flattened.
And since for a flat rotation curve, the mass is increasing linearly with radius,
the inside force is very quickly dominated by the mass outside,
which is completely unknown.

If the halo is truncated at the last HI measured radius, it can be 
highly flattened.
Paradoxically, for the same rotation curve, the dark potential of
the ``maximal halo/minimal disk'' solution is rounder than for the
``maximal disk/minimal halo'' solution, since there is then more mass in
the outer parts.
Introducing a truncation in the dark halo outside of the
last observed HI point makes
it much more flattened for a given HI thickness: for example
the flattening derived for NGC 4013 is q=0.1, for M31, q=0.2
(Becquaert, 1997).

Another method has been used to obtain the shape of dark haloes
around elliptical galaxies possessing diffuse X-ray emission
(Buote et al. 2002): haloes are triaxial, with significant
flattening. A compilation of all published results has been
recently made by Merrifield (2002), and an updated version
is reproduced in fig \ref{flat}).

Weak lensing by galaxies (or galaxy-galaxy lensing) is a new method
that can bring information on the galaxy potential at large
scale, both the 3D shape and the radial extent (see Hoekstra's
talk at this workshop). The first results
point to flattened haloes (Hoekstra et al 2002).

As for the MOND hypothesis,  Milgrom (2001) has demonstrated
that it is equivalent to assume two dark matter components,
with respect to the z-behaviour: a disk and a round halo. 
In the flaring region, the fake massive disk dominates the true one, 
and therefore the z-force is larger than the Newtonian one:
the halo should appear quite flattened, through the flaring method.
At larger radii (and larger z), the fake round halo dominates.

\section{How Far do Haloes Extend?}
\label{extension}

We know rotation curves in the outer parts of spiral galaxies only through the HI
gas kinematics.
But the HI disc is typically observable up to $R_{25}$ after which the 
neutral gaseous disc is sharply truncated, and we have no more information.

\smallskip

The HI cut-off is probably due to an ionisation of the HI by the extragalactic UV field radiation. 
Maloney (1993) describes such a process in the case of NGC 3198 and gives a profile of the 
HI decrease with a cut-off radius around the column density of 10$^{19}$ cm$^{-2}$, at 
several times the stellar disc radius. Bland-Hawthorn al. (1997) suggest 
another mechanism: the photoionization of the HI might be caused by hot and young 
inner stars. 
To understand the sharp edge in the atomic hydrogen disc,  Bland-Hawthorn al. (1997) have 
searched and detected ionised hydrogen, beyond the edge of the HI disc of NGC 253. 
More essential, the authors used the H$\alpha$ velocity to extend the rotation curve of this 
galaxy and conclude the rotation curve may fall near the HI cut-off. They even find
a hint for the expected increase-before-drop signature in the rotation curve
of a truncated disk (cf Casertano 1983). This suggests that the 
egde of the dark matter component is not far from the HI truncature. This result
is only tentative, and it is of first importance to confirm it and repeat
in other galaxies, to determine the nature and distribution of dark matter.
Evidence for the Keplerian falloff (and possibly the truncation signature)
would help to know the total mass of spiral galaxies and to test models of 3D dark 
matter structures, their flattening and radial extension.

Moreover, a truncation signature, and a fall-off at large radius would bring strong
constraints to the MOND hypothesis. 

\section{Tidal Tails}

The length, thinness and general morphology of tidal tails in 
interacting galaxies, such as the Antennae, are quite sensitive to 
the halo mass distributions in the parent galaxies  (Dubinski et al 1996).
As the mass and extent of the dark halo increase in the model galaxies, the resulting tidal 
tails become shorter, less massive, and less striking, even under the most
favorable conditions for producing tidal features. Simulations and their statistics
can then constrain the amount of dark matter around spiral galaxies, with a dark-to-visible
mass ratio less than 10.

Since these arguments encountered some controversies, mainly that the tidal tails
do not constrain the total dark matter content, but only its radial distribution,
Dubinski et al (1999) subsequently explored many different distributions and shapes.
They conclude that tidal tails formation is inhibited in 
a galaxy with a rising or flat rotation curve
dominated by the halo, unless the halo is abruptly cut off just beyond
the disk edge. On the contrary, tidal tails such those currently observed are easy to form in
galaxies with declining rotation curves, resulting either from compact, low-mass
halos or from massive disk components in low-concentration dark halos.
The galaxy models that appear to fit most of the
observational constraints are those that have disk-dominated rotation curves 
and low-concentration halos.

These findings appear to put CDM predictions in difficuty. It would be interesting to
make simulations of galaxy interactions and tidal tails formation within the
MOND hypothesis.

\begin{figure}
\resizebox{12cm}{!}{\includegraphics{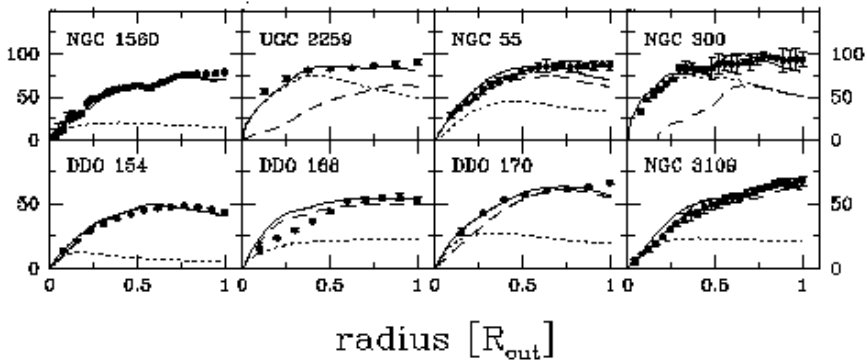}}
\caption{Rotational velocity (in km/s), versus radius,
normalised to the outer one, for dwarf galaxies, from
Hoekstra et al (2001). The HI observed rotation curve is represented by dots
with error bars, and the fit is the solid curve. The contribution of the 
stellar component is the dotted curve, and the scaled HI component is
the dashed curve (curve obtained when the HI surface
density is multiplied by a factor constant in radius).
The scale factors, to obtain the fits in each galaxy, are all gathered around 10, with a small
scatter.}
\label{has01}
\end{figure}

\section{Local Dark Matter?}

Since the pionneering study by Oort (1960) who found some dark matter
in the Milky Way disk near the Sun,
many studies were carried out before Hipparcos with contradictory
results: Bahcall (1984) finds that half of the local mass is dark,
while the conclusions of Bienaym\'e et al (1987) and Kuijken \& Gilmore (1989)
were compatible with no dark mass in the disk.
Bahcall et al (1992) quantified that locally there is 53\% more dark matter
 than visible stars. Cr\'ez\'e et al (1998) from Hipparcos data concluded
to no disk dark matter. However this result relies on
the simplifying assumptions of axisymmetry and stationarity, both
not satisfied due to the presence of a contrasted spiral structure and expected 
strong evolution in a barred galaxy.  The derived stellar
stellar density locally is 0.04 M$_\odot$/pc$^3$, while the total density 0.08 M$_\odot$/pc$^3$.
The difference is assumed to be the gas density, which is not well known.

Recently, Ibata et al. (2001) identified a stellar stream
assumed to be coming from the disruption of the Sagittarius dwarf,
through cool carbon giant stars in the Galactic halo. They argue that the orbits of these
stars in the dark matter halo of the Milky Way and the morphology of the stream put
constraints on the flattening of the halo, which they found quasi spherical.

\begin{figure}
\resizebox{12cm}{!}{\includegraphics{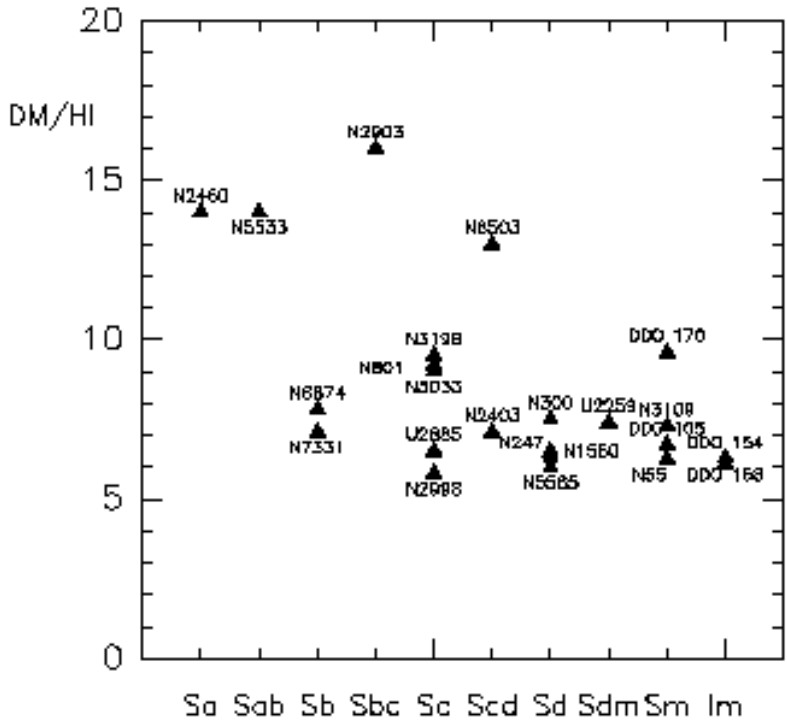}}
\caption{Dark matter to HI surface densities ratio required to
fit rotation curve of galaxies, according to their type (Combes 2000). 
The data for 23 galaxies have been taken from the compilation of Broeils (1992).}
\label{dmhi}
\end{figure}

\section{Tully-Fisher Relation}
\label{TF}

The Tully-Fisher relation, which has relatively small scatter when near infrared photometry is 
used, tells us that galaxy disks obey a scaling law (mass-to-size relation) in addition to the 
virial theorem, and to an almost constant mass-to-light ratio, over a large luminosity range. 
Since the relation involves the global velocity width of the galaxy, it is strongly weighted to 
the central parts, where all the velocity gradient is observed in general (with exception for 
dwarfs, with still rising rotation curves). The relation therefore does not tell us about the 
dark matter, which is not dominant in the central parts for bright spirals. However, it becomes
 a precious tool to detect galaxies that are dominated by dark matter in their central parts, 
where the M/L ratio becomes high.

Precisely the relation breaks down towards low luminosity galaxies, and
extreme late-type spirals (Matthews et al. 1998). Then the gas mass which becomes significant,
 has to be taken into account in the "luminosity" of the galaxy. The gas mass fraction can be 
very large, and for LSB dwarfs reach the highest levels of any known galaxy type (fg=95\%) 
(Schombert et al. 2001). The gas mass fraction is strongly correlated with
luminosity and surface brightness (McGaugh \& de Blok, 1997). Adding 
the gas ``luminosity'' to the optical luminosity is known as 
{\it baryonic correction} (Milgrom \& Braun 1988). With this correction, gas-dominated dwarf
 galaxies follow 
the same TF relation as for bright spiral galaxies. 
McGaugh et al. (2000) call this the ``baryonic TF relation''. The relation is plotted in terms 
of mass versus velocity, assuming a constant M/L ratio for the stellar component. Then, to 
compensate for the faint galaxies break, i.e. for $V_{rot} \le 90$ km s$^{-1}$, the gas mass 
is added to the stellar mass, to compute the total visible baryonic mass $M_d=M_{\star}+M_{gas}$. 
With this total mass, the TF relation
is satisfied over the entire mass range, confirming that the relation only involves
baryonic matter, which is in accordance with the MOND hypothesis.

\begin{figure}
\rotatebox{-90}{
\resizebox{11cm}{!}{\includegraphics{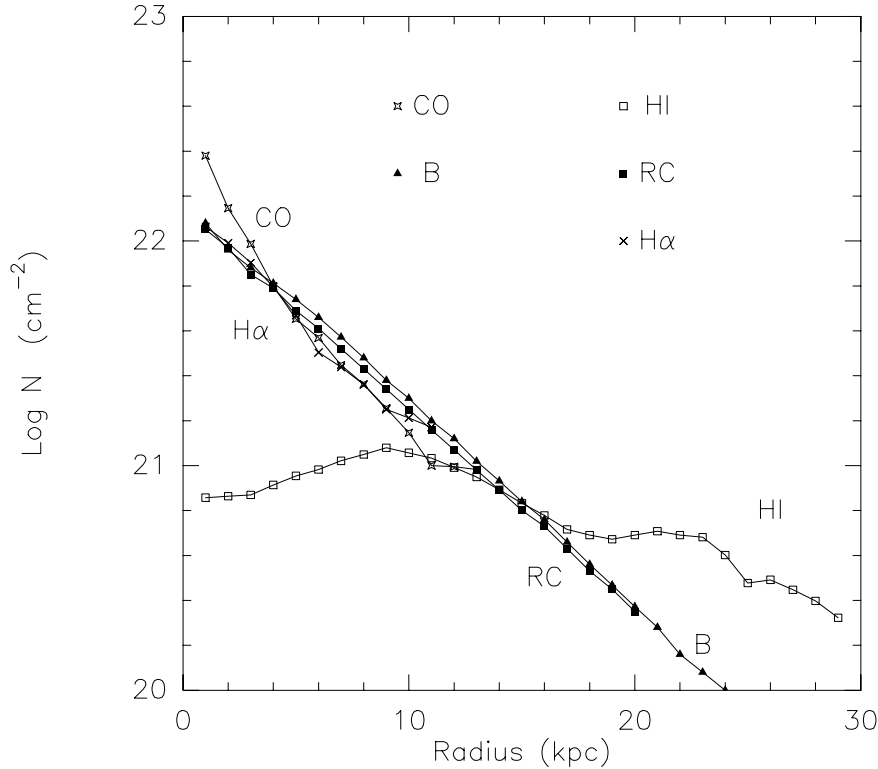}}}
\caption{ Radial distributions of various surface densities in
a typical spiral galaxy NGC 6946: H$_2$(CO) and HI column densities,
Blue, Radio-continuum and H$\alpha$ surface densities
(adapted from Tacconi \& Young, 1986).}
\label{6946}
\end{figure}

\section{Baryonic Dark Matter}
\label{BDM}

If we want to test any proposed theory of gravity, such as MOND
that replaces the non-baryonic dark matter, we still have to take 
into account the baryonic dark matter, which dominates the 
visible matter, and can change considerably the constraints on MOND,
depending on it spatial distribution.

The quantity of baryons in the Universe (and more
precisely the fraction of the critical density in baryons
$\Omega_b$) is constrained by the primordial nucleosynthesis
to be $\Omega_b =$ 0.015 h$^{-2}$, with h = H$_0$/(100 km/s/Mpc)
is the reduced Hubble constant. With h = 0.4, $\Omega_b$
is 0.09, and more generally $\Omega_b$ is between 0.01 and 0.09
(Walker et al. 1991, Smith et al. 1993), while the
visible matter corresponds to  $\Omega_* \sim $0.003 (M/L/5)
h$^{-1}$ (+ 0.006 h$^{-1.5}$ for hot gas). Therefore, most
of the baryons (90\%)  are dark. This is suported by 
the recent measurements of the CMB anisotropies (BOOMERANG, MAXIMA,
cf Jaffe et al 2000).

\subsection{Nature of the baryonic dark matter}

Since about a decade, the microlensing experiments have
accumulated lensing events by compact objects in the Milky Way
halo (Lasserre et al 2000, Alcock et al. 2001), and brown dwarfs 
are now ruled out as candidate for the baryonic dark matter.
There could be a small contribution in white dwarfs, but the bulk
of the mass has to be contributed by other candidates. These can 
only be gas now, either hot ionised gas around filaments in the 
intergalactic space (of which only a fraction $\sim$ 10$^{-4}$ is detected 
through the neutral fraction in Ly$\alpha$ absorption lines), or cold neutral
molecular gas associated with galaxies.

One model proposes to extrapolate the visible gaseous disk
towards large radii, with thickening and flaring, following the HI disk.
 The cold and dark H$_2$ component is supported by rotation,
exists only outside the optical disk, where it is required by rotation curves
(Pfenniger et al 1994, Pfenniger \& Combes 1994). The gas is stabilised
through a constantly evolving fractal structure, experiencing
Jeans instabilities at all scales, in thermal equilibrium with the
cosmic background radiation at T = 2.7 (1+z) K.

Other models distribute the dark molecular gas in a spherical or
 flattened halo, with no hole within the optical disk. The molecular
gas is not so cold, and is associated with clusters of brown dwarfs
or MACHOS (de Paolis et al. 1995, Gerhard \& Silk 1996, Kalberla
et al. 1999).

\subsection{Coupling between HI gas and dark matter}

In the first model, the HI gas can be considered as a tracer of the
dark baryons,
the interface between the molecular clumps and the extra-galactic radiation field.
Beyond the HI disk, there could be an ionization front, and the interface might
become ionized hydrogen. In this context, there should exist a distribution
correlation between the dark matter and the HI gas. This is indeed the
case, as already remarked by Bosma (1981), Broeils (1992) or Freeman (1993):
there is a constant ratio between the surface density of dark matter, as
deduced from the rotation curves, and the HI surface density,
$\Sigma_{DM} / \Sigma_{HI}$ =  7-10 (cf figure \ref{has01}, extracted 
from Hoekstra et al. 2001). This
ratio is constant with radius in a given galaxy, and varies
slightly from galaxy to galaxy, being larger in early-types
(figure \ref{dmhi}).
However, the dark matter does not dominate the mass in the latter, and
therefore the estimate of its contribution is more uncertain.
The correlation is the most striking in dwarf galaxies, which are
dominated by dark matter (figure \ref{has01}).

\begin{figure}
\resizebox{12cm}{!}{\includegraphics{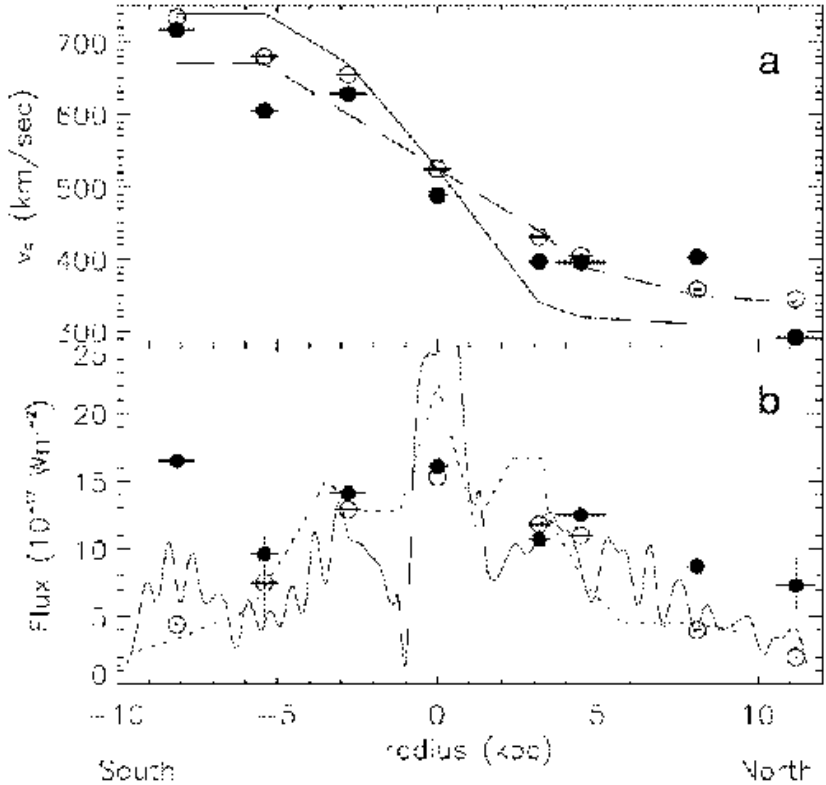}}
\caption{ Major axis profiles in the edge-on galaxy NGC 891 of
the two fondamental rotation lines of H$_2$: 
S(0)(filled circles) and S(1) (open circles), compared to the CO
profile (full line); a) velocity profiles
b) integrated line strengths (from Valentijn \& van der Werf 1999).} 
\label{val99}
\end{figure}

\subsection{H$_2$/HI ratio and its radial variation}

The differences between HI and H$_2$ (traced 
by CO line emission) radial
distributions in galaxies is striking (cf figure \ref{6946}).
While all components related to star formation,
the blue luminosity from stars, the H$\alpha$ (gas ionised
by young stars), the radio-continuum (synchrotron related
to supernovae), and even the CO distribution, follow
an exponential distribution, the HI gas alone is
extending much beyond the ``optical'' disk, sometimes
in average by a factor 2 to 4 (R$_{HI}$ = 2-4 R$_{opt}$).

In fact, the true H$_2$ radial distribution is not known,
since the CO emission is not a good tracer, especially
because it depends on metallicity (may be in a non-linear way);
it is well known that the metallicity decreases exponentially
with radius in typical spirals.  The CO-emission 
exponential fall off has therefore to be corrected to
deduce the true H$_2$ distribution. Given that the 
H$_2$/HI surface density ratio is larger than 10 in the center,
it is not impossible that cold H$_2$ exists in such proportions 
in the outer parts
as to account for rotation curves in spiral galaxies.

\begin{figure}
\resizebox{12cm}{!}{\includegraphics{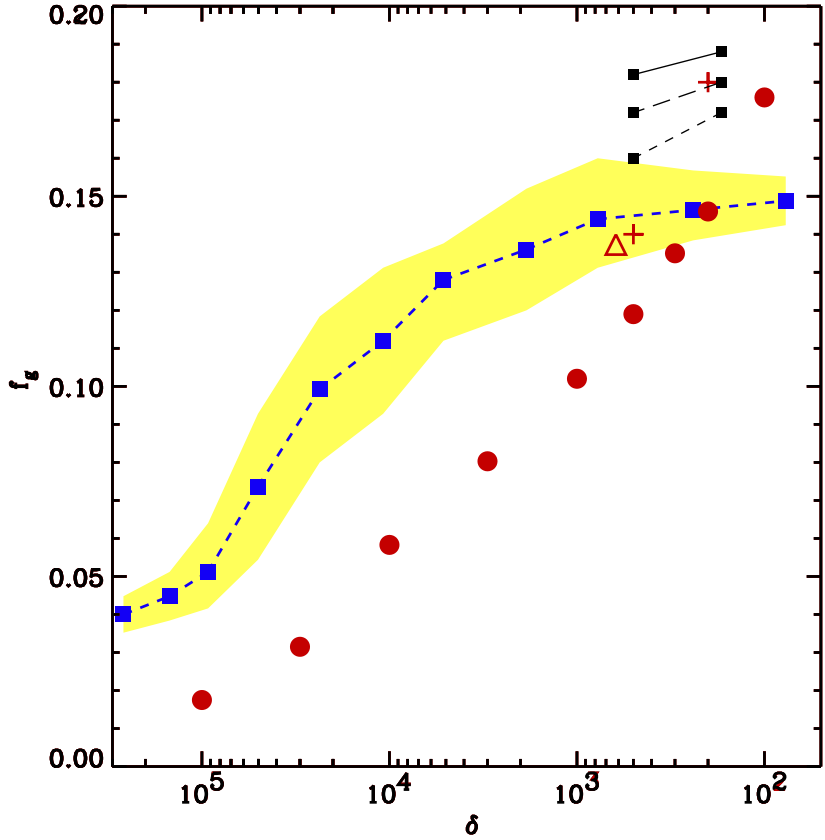}}
\caption{Radial distribution of the hot gas fraction f$_g$ in clusters of galaxies:
the abscissa $\delta$ is the average density inside the radius r, normalised
to the critical density. The filled circles, empty triangle and crosses are the data,
while the squares connected by a dashed line are the theoretical predictions
(cosmological simulations from several groups), 
with an assumed baryon fraction of f$_b = \Omega_b/\Omega_m$ =0.16; 
other model variants (with or without stellar feedback, supernovae winds)
are shown with f$_b = \Omega_b/\Omega_m$ =0.20 (small squares), from Sadat \& Blanchard (2001)}
\label{sadat01}
\end{figure}

\subsection{Warm H$_2$ as a tracer?}

H$_2$ is a symmetric molecule, and does not radiate at low temperature
(the first accessible level with quadrupole transitions is at 500K 
above the fundamental). Among the various methods to try
to detect it indirectly (e.g. Combes \& Pfenniger 1997), one of the
most favorable is to observe the fundamental pure rotational lines
S(0) at 28 $\mu$m and S(1) at 17 $\mu$m. Gas must be warm ($\sim$ 100K)
to have some significant emission, but some warm gas is always expected
to be present in a turbulent fractal medium, where small clumps enter
in collisions frequently. Slow shocks can then heat some small fraction
of the gas, such that emission in the first rotational lines is detectable.  

Already ISO observations of
these lowest pure rotational lines of H$_2$ in NGC 891 have brought clues
for the existence of large quantities of H$_2$ in galaxies
(Valentijn \& van der Werf 1999). The H$_2$ emission is not decreasing
steeply with radius, as is the CO integrated emission
(see fig \ref{val99}). The warm H$_2$ gas is well mixed with the CO
gas, according to its kinematics. The linewidths suggest that the 
more fundamental line S(0) is more extended than S(1). Although the observations
stop at the end of the optical disk, it seems possible to detect the emission
in the outer parts. 

These observations should be pursued
in external galaxies, at much further radius than was possible with ISO.
If the presence of large dark gas is confirmed in galaxies, it will change
drastically the MOND predictions, since this dark baryons have not the
same radial distribution than the visible matter.

\subsection{Distribution of baryonic matter in galaxy clusters}

In galaxy clusters, the hot gas detected in X-rays dominate the visible
mass. Depending on the cluster, most or all of the baryons have become luminous,
since the visible baryonic mass fraction is representative of the whole
universe f$_b = \Omega_b/\Omega_m \sim$ 0.15. A striking feature
is the radial distribution of visible and dark masses, which is now reversed
at those scales. Indeed, if the dark matter fraction is increasing with radius
in galaxies, it is decreasing with radius in clusters (cf fig \ref{sadat01}).
This result was already emphasized by David et al (1995), and
has been confirmed and precised since then (Ettori \& Fabian 1999, Sadat \& Blanchard 2001).
The gas mass fraction ranges from 10 to 25\%, and varies from cluster to cluster.
These variations  may be explained if the dark matter has a significant baryonic component. 
In those clusters where no significant amount of baryonic dark matter remain, it is
quite difficult to maintain a MOND interpretation of the non-baryonic dark matter,
which is more concentrated than the visible matter. Already Milgrom (1998) acknowledged
that the MOND hypothesis was not able to account for clusters cores, except
in the presence of a dominant baryonic dark component in the center.

\section{Conclusions}

Dark haloes at galactic scales are now constrained by more precise rotation
curves. Bright spiral galaxies are not dominated by dark matter in their optical
disks. The dark-matter/visible mass ratio is a function of luminosity and surface
brightness. Dwarf and LSB galaxies are the best laboratories for dark matter
studies since they are dominated by unseen matter down to their central regions.
The derived radial profile of dark matter is not centrally concentrated and
presents no cusp as predicted in the CDM scenario.

The 3D shape of haloes is still badly constrained. 
Polar rings are often self-gravitating
and there are some clues that their potential is flattened along the polar plane.
The HI plane flaring method depends strongly on the assumed truncation radius of
the dark matter component.

Observations have shown however that haloes are oblate and galaxy
potential axisymmetric in their planes.

Statistical galaxy-galaxy lensing might bring some progress in the determination
of shape and radial extension of dark matter haloes.

The formation of tidal tails in galaxy interactions is a good test of the
shape of their potential. Simulations have shown that only galaxies dominated 
by their visible matter, or with their halo truncated outside their optical
disk, were able to form tails corresponding to observations. Such simulations
should be explored within the MOND hypothesis.

Most baryons are dark, according to primordial nucleosynthesis and CMB
anisotropies. These baryons could be present in the form of cold molecular
clouds in the outer parts of galaxies, with a H$_2$/HI surface density ratio 
of about 10, as suggested by rotation curves. This reservoir of gas in the outer parts
account for galaxy evolution, that requires fresh replenishment of gas for star
formation, and explains the evolution of morphological types along the Hubble sequence:
late-types have a much larger proportion of dark matter than early-types, while 
secular evolution (through bars and spirals), and interactions/mergers tend
to progressively transform late-type galaxies in early-type ones.

In galaxy clusters, the baryonic matter is almost all visible in the form
of hot X-ray gas. The distribution of the dark with respect to visible matter,
which increases with radius at galactic scales, and then decreases with radius
at cluster scale, might raise strong constraints in all modified gravity/dynamics
theories. For all these scenarii, the spatial distribution of
baryonic dark matter is a fundamental element to consider.


\end{document}